\documentclass[12pt]{article}
\usepackage[applemac]{inputenc}
\usepackage{epsfig}
\begin{document}

\date{\today}

\title{Removing the gauge parameter dependence of the effective potential  by a field redefinition \footnote{Preprint Numbers: CP3-Origins-2014-23 DNRF90 and DIAS-2014-23}}

\author{N. K. Nielsen
\footnote{email: nkn@cp3-origins.net}\\
Center of Cosmology and Particle Physics Phenomenology (CP3-Origins),\\
University of Southern Denmark, \\
Odense, Denmark}

\maketitle

\begin{abstract}
The gauge parameter dependence of the effective potential is determined by partial differential equations involving also the Higgs boson field expectation value. Solving these equations  by the method of characteristics leads to  elimination  of the gauge parameter dependence of the effective potential. The construction is carried out  in the case of the standard model of electroweak unification for the renormalization group improved effective potential up to the next-leading logarithmic order.\\ 
 {\em PACS  numbers 11.15.-q, 12.15.-y, 12.15.Lk, 14.70.-e.}\\
\end{abstract}

\newpage

\section{Introduction} 

With the discovery of the Higgs boson \cite{ATLAS}, \cite{CMS} the theoretical determination of its lower mass bound by  means of the  renormalization group method improved effective potential \cite{Cabibbo}, \cite{Casas} has gained new prominence \cite{Degrassi}. The stability analysis of \cite{Cabibbo}, \cite{Casas}, \cite{Degrassi} takes place in a specific gauge (the Landau gauge). 

It was pointed out some time ago by  Loinaz and Willey \cite{Willey} that the gauge dependence of the effective potential could make the lower bound of the Higgs boson mass gauge dependent. They used an Abelian truncation of the standard model of electroweak unification, which in the bosonic sector is identical to the abelian Higgs model,  to make their point, working to one-loop and leading logarithmic  order.  A recent study by Andreassen \cite{Andreassen} gives a very useful review of this difficulty. It is a problem of both phenomenological and conceptual importance, considering the fact that the observed Higgs mass is very close to the stability limit of the standard model of electroweak unification.  It has  been shown that the same problem arises in the context of Higgs inflation \cite{Krauss}. Quite recently the problem has been dealt with in connection with the so-called instability scale \cite{Luzio}.

It will be shown in this paper that the gauge dependence of the effective potential can be removed by solving a set of homogeneous first order partial differential equation obeyed by the effective potential and involving the gauge parameters and the Higgs boson field expectation value. The validity of these equations
 was established many years ago  by the author \cite{NKN}  and independently by Fukuda and Kugo \cite{Fukuda} (see also \cite{Aitchison}, \cite{Sibold}, \cite{Piguet}). Solving the differential equations  leads to a field redefinition that eliminates the gauge dependence from the effective potential. The analysis is carried out in the context of the $SU(2)\otimes U(1)$ standard model of electroweak unification for the renormalization group improved effective potential at the leading and   next-leading logarithmic order.

Suppressing in the effective potential $V$ all variables except the Higgs field expectation value $\phi$ and a gauge parameter $\xi $, the partial differential equations found in \cite{NKN} and \cite{Fukuda} have the form:
\begin{equation}
(\xi\frac{\partial}{\partial \xi}+C(\phi, \xi)\frac{\partial}{\partial \phi})V(\phi, \xi)=0
\label{identity}
\end{equation}
with $C(\phi, \xi)$ a calculable function.   (\ref{identity})  can, as pointed out in \cite{NKN} and \cite{Piguet},  be solved by the method of characteristics. It is an obvious consequence of (\ref{identity}) that the potential $V$ is unchanged if  the gauge parameter $\xi$  and the field $\phi$ are subject to the following  changes:
\begin{equation}
\xi \rightarrow \xi +\Delta \xi, \phi \rightarrow \phi+C(\phi, \xi)\frac{\Delta \xi}{\xi}
\label{stormare}
\end{equation}
with $\Delta \xi$ infinitesimal.
Taking this argument one step further, one can introduce a new field variable $\Phi(\phi, \xi, \xi_0)$ that is a solution of the partial differential equation:
\begin{equation}
(\xi\frac{\partial}{\partial \xi}+C(\phi, \xi)\frac{\partial}{\partial \phi})\Phi (\phi, \xi, \xi_0)=0
\label{Fukuda}
\end{equation}
with the boundary condition:
\begin{equation}
\Phi(\phi, \xi, \xi_0)\mid _{\xi=\xi_0}=\phi
\label{Kugo}
\end{equation}
with $\xi_0$ a specific value of the gauge parameter $\xi$.  Expressing the effective potential in terms of the new field variable one can now eliminate the dependence  of the effective potential on the gauge parameter  $\xi $ altogether since $\Phi(\phi, \xi, \xi_0)$ is invariant under (\ref{stormare}) and thus is equal to the field variable at $\xi=\xi_0$, and  (\ref{Fukuda}) and (\ref{Kugo}) imply:
\begin{equation}
V(\phi, \xi)=V(\Phi(\phi, \xi, \xi_0), \xi_0).
\label{Kyriakopoulos}
\end{equation}

What has been achieved by this construction is to show that  the effective potential at an arbitrary value $\xi$ of the gauge parameter is equal to the potential at a  preferrred value of the gauge parameter $\xi_0$ by introducing a 
mapping through the solution of (\ref{Fukuda})-(\ref{Kugo}):
\begin{equation}
\phi \rightarrow \Phi(\phi, \xi, \xi_0).
\label{mapping}
\end{equation}
The inverse mapping is obviously obtained by interchanging $\xi$ and $\xi_0$ in $\Phi(\phi, \xi, \xi_0)$. As mentioned above the stability analysis giving the  Higgs boson mass lower bound \cite{Cabibbo}, \cite{Casas}, \cite{Degrassi} takes place in the Landau gauge, and the construction described above makes it possible to extend this analysis to arbitrary gauges; this is the problem raised in \cite{Willey}, \cite{Andreassen},  \cite{Krauss}, \cite{Luzio}. Our approach to this problem should be contrasted to that of \cite{Boyanovsky} and \cite{Lin}, who suggested using instead  a gauge-independent effective  potential constructed either by a Hamiltonian method or by the Vilkovisky-DeWitt formalism \cite{Vilkovisky}, working in an Abelian model similar to the model used in \cite{Willey} to leading logarithmic order.

The Higgs boson mass defined by the propagator pole  is independent of the gauge parameters at all values of the quartic coupling $\lambda$ entering the tree approximation potential term $\frac 14 \lambda \phi^4$, provided one takes into account that the field value at the electroweak potential minimum is gauge parameter dependent; this statement was proven  in \cite{NKN} sec.4 and also verified in \cite{Aitchison}  (these proofs were carried out for the Abelian Higgs model but are easily extended to the standard model of electroweak unification). However, use of the gauge-dependent effective potential is necessary in order to determine the value of $\lambda$ that leads to the lower bound on the Higgs mass such  that  the electroweak vacuum is stable. Extension of the analysis of this problem  in  \cite{Cabibbo}, \cite{Casas}, \cite{Degrassi} beyond the Landau gauge can be carried out by (\ref{mapping}) or its inverse, which relate the field value at one set of values  of the gauge parameters to the field value at another set of values of the gauge parameters, such that the value of the effective potential is unchanged.

In order to make (\ref{Kyriakopoulos}) useful one has to provide a solution of (\ref{Fukuda})-(\ref{Kugo}). Usually the effective potential is found in the context of some approximation scheme, where it is obtained order by order. It is rather straightforward to find approximate solutions of  (\ref{Fukuda})-(\ref{Kugo}) in the context of the loop expansion. In   connection with the lower bound of the Higgs boson mass one is, however, dealing with the renormalization group improved effective potential, which involves a resummation of the effective potential at infinite loop order. Explicit  solutions of (\ref{Fukuda})-(\ref{Kugo}) are, as mentioned above, obtained in this paper in the leading and next-leading logarithmic order for the standard model of electroweak unification, though there seems no reason why it should not be extended to arbitrary orders in this expansion. Only the bosonic part of the  standard model of electroweak unification is considered here since this is where complications involving gauge parameter dependence occurs; it is trivial to include also the fermionic part.

The outlay of this paper is the following: Sec. 2 reviews ingredients necessary for the construction, where in sec. 2.1  material on the standard model of electroweak unification is collected and in sec. 2.2 the one-loop effective potential is briefly reviewed, while  sec. 2.3 deals with renormalization group improvement of the effective potential. Here the leading  and next-leading approximations of the renormalization group solutions are defined, and it is is noticed that different solutions of the renormalization group equation of the effective potential may have different orderings in this approximation scheme. In sec.3   the leading and next-leading logarithmic approximations of the effective potential are constructed, and (\ref{identity}) is verified  and the relevant functions $C(\phi, \xi)$ are determined. Finally in sec. 4 these functions are used to solve (\ref{Fukuda})-(\ref{Kugo}) and to check (\ref{Kyriakopoulos}), such that the  field redefinition alluded to in the title of the paper is achieved. An appendix lists the renormalization group functions used.

\section{Preliminaries}

\subsection{$SU(2)\otimes U(1)$ electroweak theory}

With $\phi$ the expectation value of the Higgs boson field the effective potential of the $SU(2)\otimes U(1)$ standard model of electroweak unification is in the tree approximation:
\begin{equation}
V^{[0]}(\phi)=\frac 12\mu^2\phi^2+\frac 14\lambda \phi^4, \ \mu^2<0, \ \lambda >0.
\label{hedeprinsesse}
\end{equation}

The  theory contains the gauge fields $W^{\pm}_\mu, W^3_\mu$ and  $B_\mu$ and in a renormalizable gauge the Goldstone boson fields $\chi^\pm$ and $\chi^3$. A possible and rather general   gauge fixing  term of the Lagrangian is:
\begin{eqnarray}&&
\mathcal{L}_{\rm gf}  
=-\frac {1}{\xi} (\partial _\mu W^{+,\mu}-\frac{gu}{2}\chi
^+)(\partial _\nu W^{-,\nu}-\frac{gu}{2}\chi
^-) 
\nonumber\\&&
-\frac{1}{2\xi}( \partial_\mu W^{3\mu}
-\frac{gu}{2}\chi ^3)^2-\frac{1}{2\xi'}(\partial _\mu
B^\mu+\frac{g'u'}{2}\chi ^3)^2
\label{hakonjarl}
\end{eqnarray}
with $\xi, \xi ', u$ and $u'$ gauge parameters and where $g$ and $g'$ are the $SU(2)$ and $U(1)$ coupling constants. This leads   to the following ghost Lagrangian in the bilinear approximation:
\begin{eqnarray}&&
{\cal L}_{\rm FP} = - \bar{c}^+(\partial
-\frac{g^2u\phi}{4})c^+
- \bar{c}^-(\partial
^2-\frac{g^2u\phi }{4})c^-- \bar{c}^3\partial ^2c^3- \bar{c}'\partial ^2c'
\nonumber\\&&
+\frac{\phi}{4}(gu\bar c^3-g'u'\bar c')(gc^3-g'c')
\label{conelvito}
\end{eqnarray}
with $c^+, \bar c^+, c^-, \bar c-, c^3, \bar c^3, c'$ and $\bar c'$ Faddeev-Popov ghost fields.
The gauge parameters $u$ and $u'$ should be treated as independent variables in the context of the renormalization group since the gauge fixing Lagrangian is not renormalized \cite{Zinn} and the combinations $gu\chi^\pm$ and $gu\chi^3$ are thus renormalized by the same multiplicative renormalization as $W^\pm_\mu$ and $W^3_\mu$, while $g'u'\chi^3$ is renormalized like $B_\mu$. 
At general values of $u$ and $u'$ the vector and Goldstone boson fields mix in the bilinear approximation of the Lagrangian.

Consider first the charged vector fields. 
They  are St\"{u}ckelberg decomposed:
\begin{equation}
W_\mu ^\pm= W_{tr,\mu}
^\pm+\frac{1}{\sqrt{\partial^2} } 
\partial_\mu\omega  ^\pm
\label{eq:Stuckelberg}
\end{equation}
where:
\begin{equation}
\partial_\mu W_{tr}^{\pm, \mu } =0.
\end{equation}
In momentum space with $k$ a momentum variable  this gives the following Lagrangian bilinear in the charged vector and Goldstone boson fields:
\begin{equation}
{\cal L}=-W_{tr}^{+\mu }(-k)(k^2-\frac{g^2\phi ^2}{4}) W^-_{tr,\mu}(k)
-(\chi ^+(-k), \omega ^+(-k)) 
\left(\begin{array}{ll}
a_{11}(k)&a_{12}(k)\\a_{12}(k)& a_{22}(k)\end{array}\right )
\left(\begin{array}{l}
\chi ^-(k)\\\omega ^-(k)\end{array}\right )
\label{eliwallach}
\end{equation}
where:
\begin{equation}
a_{11}(k)=
-k^2+\mu^2+\lambda \phi^2+\frac{g^2u^2}{4\xi}, \
a_{22}(k)=- \frac{k^2}{\xi} +M_W^2, \
a_{12}(k)=-\frac g2(\phi+\frac{u}{\xi})\sqrt{-k^2}
\end{equation}
with $M_W^2=\frac{g^2\phi^2}{4}$ and with the determinant:
\begin{equation}
{\rm Det}\left(\begin{array}{ll}
a_{11}(k)&a_{12}(k)\\a_{12}(k)&a_{22}(k)\end{array}\right
)
=\frac {1}{\xi}(k^2-k_{+,W}^2)(k^2-k_{-, W}^2)
\label{sonjadantziger}
\end{equation}
where $k_{\pm, W}$ are defined by:
\begin{equation}
k_{\pm, W}^2 = \frac 12\big (\mu^2 +\lambda \phi^2 - \frac 12 g^2 u \phi
\big ) \pm \frac 12\sqrt{\big (\mu^2+\lambda \phi^2 \big)\big
(\mu^2+\lambda \phi^2 -  g^2 \phi(u+\xi\phi) \big ) }.
\label{jeanette}
\end{equation}

The neutral vector fields  are also St\"{u}ckelberg decomposed:
\begin{equation}
W_\mu ^3= W_{tr,\mu}
^3+\frac{1}{\sqrt{\partial^2} } 
\partial_\mu\omega  ^3,  B_\mu = B_{tr,\mu}
+\frac{1}{\sqrt{\partial^2} } 
\partial_\mu\omega  '
\label{II:Stuckelberg}
\end{equation}
where:
\begin{equation}
\partial_\mu W_{tr}^{3, \mu } =\partial_\mu B_{tr}^{ \mu } =0
\end{equation}
leading to 
the quadratic Lagrangian is in momentum space:
\begin{eqnarray}&&
{\cal L}=\frac 12\bigg(\bigg (W_{\rm {tr}, \mu}^3(-k), B_{{\rm tr},\mu }(-k)\big ) \left (\begin{array}{ll}
-k^2+\frac{g^2\phi ^2}{4}&-\frac{gg'\phi^2}{4}\\-\frac{gg'\phi^2}{4}&-k^2+\frac{g'^2\phi^2}{4}\end{array}\right ) \left (\begin{array}{l}W_{\rm {tr}}^{3\mu}(k)\\B_{\rm {tr}}^\mu(k)\end{array}\right )
\nonumber\\&&
-(\chi ^3(-k), \omega ^3(-k), \omega '(-k)) 
\left(\begin{array}{lll}
b_{11}(k)&b_{12}(k)&b_{13}(k)\\b_{12}(k)& b_{22}(k)&b_{23}(k)\\  b_{13}(k)&b_{23}(k)&b_{33}(k)\end{array}\right )
\left(\begin{array}{l}
\chi ^3(k)\\\omega ^3(k)\\\omega '(k)\end{array}\right )\bigg ),
\label{stuartlittle}
\end{eqnarray}
with:
\begin{equation} 
b_{11}(k)=
-k^2+\mu^2+\lambda\phi^2 + \frac{g^2u^2}{4\xi }+\frac{g'^2u'^2}{4\xi'}
\label{lilopulver}
\end{equation}
\begin{equation}
b_{12}(k)=-\frac g2(\phi+\frac{u}{\xi})\sqrt{-k^2}, \ \
b_{13}(k)=\frac {g'}{2}(\phi+\frac{u'}{\xi'})\sqrt{-k^2},
\label{bond}
\end{equation}
\begin{equation}
b_{22}(k)=-\frac{k^2}{\xi} + \frac{g^2
\phi^2}{4}, \ \
b_{33}(k)=-\frac{k^2}{\xi '}+\frac{g'^2\phi^2}{4},
b_{23}(k)=-\frac{gg'\phi^2}{4}
\label{blofeld}
\end{equation}
with  the determinant:
\begin{eqnarray}&&
{\rm Det }\left(\begin{array}{lll}
b_{11}(k)&b_{12}(k)&b_{13}(k)\\b_{12}(k)& b_{22}(k)&b_{23}(k)\\  b_{13}(k)&b_{23}(k)&b_{33}(k)\end{array}\right )
=-\frac{1}{\xi \xi '}k^2(k^2-k_{+, Z}^2)(k^2-k_{-, Z}^2).
\label{plavalaguna}
\end{eqnarray}
Here was introduced:
\begin{eqnarray}&&
k^2_{\pm, Z}=\frac 12 (\mu^2+\lambda \phi^2-\frac 12 (g^2+g'^2)u_Z\phi)
\nonumber\\&&
\pm \frac 12\bigg((\mu^2+\lambda \phi^2)(\mu^2+\lambda \phi^2-(g^2+g'^2)\phi(u_Z+\xi_Z\phi)\bigg)^{\frac 12}
\label{ludmilla}
\end{eqnarray}
cf. (\ref{jeanette}), with the definitions:
\begin{equation}
\xi _Z= \frac{\xi g^2+\xi'g'^2}{g^2+g'^2}, \ u_Z=\frac{g^2u+g'^2u'}{g^2+g'^2}.
\label{hornhyl}
\end{equation}

\subsection{One-loop  effective potential}

In one-loop order the effective potential corresponding to the gauge fixing  (\ref{hakonjarl}) can be seen from \cite{Casas}-\cite{Luzio}, using also  (\ref{conelvito}), (\ref{eliwallach}), (\ref{sonjadantziger}), (\ref{stuartlittle}) and (\ref{plavalaguna}).
It has in $D=4-\epsilon$ dimensions the following contribution
from the Higgs field:
\begin{equation}
V_H^{[1]}= -\frac{i}{2} \int \frac{d^Dk}{(2\pi)^D} \log \frac{k^2-\mu ^2-3\lambda\phi^2+ i\epsilon}{k^2+i\epsilon}\simeq -\frac{9\lambda ^2\phi^4}{64\pi^2}(\frac 2\epsilon-\log \frac{3\lambda\phi^2}{M^2}+\frac 32)
\label{capuleti}
\end{equation}
where in the last step the asymptotic part at large valus of $\phi$ was kept.
The contribution from  the $W^\pm$-field with associated Goldstone  bosons and ghosts  is by means of (\ref{eliwallach}) and (\ref{sonjadantziger}):
\begin{eqnarray}&&
V_{W^\pm}^{[1]}= -i \int \frac{d^Dk}{(2\pi)^D}\bigg((D-1)
\log \frac{k^2-M_W^2+ i\epsilon}{k^2+i\epsilon}
+\log \frac{(k^2-k_{+,W}^2+i\epsilon)(k^2-k_{-,W}^2+i\epsilon)}{(k^2+i\epsilon )^2}
\nonumber\\&&-2\log
\frac{k^2+\frac {g^2u\phi}{4}+ i\epsilon}{k^2+i\epsilon}\bigg )
\nonumber\\&&
\simeq  -\frac{3M_W^4}{32\pi^2}(\frac 2\epsilon-\log \frac{M_W^2}{M^2}+\frac 56)-\frac{\lambda (\lambda-\frac 12\xi g^2)\phi^4}{32\pi^2}(\frac 2\epsilon-\frac 12\log \frac{\frac 14\xi g^2\lambda\phi^4}{M^4}+\frac 32)
\nonumber\\&& 
+\frac{\lambda\phi^4}{64\pi^2}\sqrt{\lambda(\lambda-\xi g^2)}\log \frac{\lambda+\sqrt{\lambda(\lambda-\xi g^2)}}{\lambda-\sqrt{\lambda(\lambda-\xi g^2)}}
\label{monteschi}
\end{eqnarray}
where in the last step again only the asymptotic part at large values of $\phi$ was kept, with the gauge parameter $u$ kept fixed.
Also we get from the $A-$ and $Z^0$-fields with associated Goldstone  boson and ghosts by (\ref{stuartlittle}) and (\ref{plavalaguna}):
\begin{eqnarray}&&
V_{Z^0}^{[1]}= -\frac{i}{2} \int \frac{d^Dk}{(2\pi)^D}\bigg((D-1)
\log \frac{k^2-M_Z^2+ i\epsilon}{k^2+i\epsilon}
+\log \frac{(k^2-k_{+,Z}^2+i\epsilon)(k^2-k_{-,Z}^2+i\epsilon)}{(k^2+i\epsilon)^2}
\nonumber\\&&-2\log
\frac{k^2+\frac {(g^2+g'^2)u_Z\phi}{4}+ i\epsilon}{k^2+i\epsilon}\bigg )
\nonumber\\&&
\simeq  -\frac{3M_Z^4}{64\pi^2}(\frac 2\epsilon-\log \frac{M_Z^2}{M^2}+\frac 56)-\frac{\lambda (\lambda-\frac 12\xi_Z(g^2+g'^2)\phi^4}{64\pi^2}(\frac 2\epsilon-\frac 12\log \frac{\frac 14\xi_Z(g^2+g'^2)\lambda\phi^4}{M^4}+\frac 32)
\nonumber\\&& 
+\frac{\lambda\phi^4}{128\pi^2}\sqrt{\lambda(\lambda-\xi_Z(g^2+g'^2))}
\log \frac{\lambda+\sqrt{\lambda(\lambda-\xi_Z(g^2+g'^2)})}{\lambda-\sqrt{\lambda(\lambda-\xi_Z(g^2+g'^2)})}
\label{montague}
\end{eqnarray}
where again only the asymptotic part, with $\phi>>u, u'$,  was kept, and with $M_Z^2=\frac{(g^2+g'^2)\phi^2}{4}$.

The infinite parts of (\ref{capuleti}),  (\ref{monteschi}) and  (\ref{montague}) 
are cancelled by the standard one-loop counterterms, and collecting  the asymptotic part one gets, with $\beta^{[1]}_\lambda$ and $\gamma^{[1]}_\phi$  one-loop renormalization groop functions (see (\ref{annyschlemm})):
\begin{equation}
V^{[1]}_{\rm ren}\simeq \frac 14(\beta^{[1]}_\lambda -4\lambda \gamma^{[1]}_\phi)\phi^4\log \frac \phi M+\frac 14\Delta \lambda \phi^4
\label{mazanderan}
\end{equation}
where:
\begin{eqnarray}&&
\Delta \lambda \simeq \frac{1}{16\pi^2}
\bigg(\frac 38g^4(\log (\frac{g^2}{4})-\frac 56)
+\frac{3}{16}(g^2+g'^2)^2(\log(\frac{g^2+g'^2}{4})-\frac 56)
\nonumber\\&&+9\lambda^2(\log(3\lambda)-\frac 32)
+2\lambda(\lambda -\frac 12\xi g^2)(\frac 12\log (\frac 14\xi g^2\lambda)-\frac 32)
\nonumber\\&&
+\lambda(\lambda -\frac 12\xi_Z(g^2+g'^2))(\frac 12\log ( \frac 14\xi_Z(g^2+g'^2)\lambda)-\frac 32)
\nonumber\\&&
+\lambda\sqrt{\lambda(\lambda-\xi g^2)}\log \frac{\lambda+\sqrt{\lambda(\lambda-\xi g^2)}}{\lambda-\sqrt{\lambda(\lambda-\xi g^2)}}
\nonumber\\&&
+\frac \lambda 2\sqrt{\lambda(\lambda-\xi_Z(g^2+g'^2))}
\log \frac{\lambda+\sqrt{\lambda(\lambda-\xi_Z(g^2+g'^2)})}{\lambda-\sqrt{\lambda(\lambda-\xi_Z(g^2+g'^2)})}
\bigg ).
\label{ignoramus}
\end{eqnarray}
The important  point concerning this  expression is that it has a nontrivial dependence on the gauge parameters $\xi$ and $\xi'$. Its value at $\xi=\xi_0$ and $\xi'=\xi'_0$, with $\xi_0$ and $\xi'_0$ specific values of the gauge parameters, is denoted $\Delta \lambda _0$.

\subsection{Renormalization group improved effective potential}

The  effective potential  is a solution of the renormalization group equation:
\begin{equation}
(M\frac{\partial}{\partial M}+\sum_i\beta _{g_i}\frac{\partial}{\partial g_i}-\gamma _\phi \phi \frac{\partial}{\partial \phi})V(\phi, M, g_i)=0
\label{oiseauxdefeu}
\end{equation}
where $M$ denotes the renormalization scale and $g_i$ are coupling constants and dimensionless gauge fixing parameters  (we only consider the renormalization group  asymptotically such that   the dependence of the effective potential on $\mu^2$ and the gauge parameters $u, u'$ can be disregarded).
Ford, Jones, Stephenson and Einhorn \cite{Ford} have found the following solution
of this equation:
\begin{equation}
V(\phi, M, g _i)=V(\phi(t), M(t), g _i (t))
\label{FJSE}
\end{equation}
with:
\begin{equation}
\phi(t)=\phi \eta (t), \ \eta(t)=\exp(-\int
_0^tdt'\gamma _\phi(g_i (t'))), \
M(t)=e^tM
\label{rykarensdal}
\end{equation}
and:
\begin{equation}
\frac{dg_i (t)}{dt}=\beta _i(g_j (t)).
\label{carlotta}
\end{equation}
In (\ref{FJSE}) one conveniently chooses $\phi(t)=M(t)$, which  by (\ref{rykarensdal}) is the same as:
\begin{equation}
 t=  \log \frac{\phi }{ M}-\int _0^tdt'\gamma _\phi (t')
 \label{nickelcoin}
\end{equation}
This choice of $t$ makes terms involving logarithmic factors $$\log \frac{\phi(t)}{M(t)}$$ disappear on the right-hand side of (\ref{FJSE}).

Another solution of  (\ref{oiseauxdefeu}) was obtained by Coleman and Weinberg \cite{Coleman}.
From dimensional considerations follows:
\begin{equation}
(M\frac{\partial}{\partial M}+\phi \frac{\partial }{\partial \phi})V=4V
\end{equation}
that is combined with (\ref{oiseauxdefeu}) which
 becomes by elemination of $\phi$:
\begin{equation}
(M\frac{\partial}{\partial M}+\bar \beta_i\frac{\partial}{\partial g_i}
-4\bar \gamma_\phi)V(\phi, M, g_i)=0
\end{equation}
with $\bar \beta_i=\frac{\beta_i}{1+\gamma_\phi}, \bar \gamma _\phi=\frac{\gamma_\phi}{1+\gamma_\phi}$ and with solution:
\begin{equation}
V(\phi, M, g _i)=\bar \eta(\bar t)^4V(\phi, M(\bar t), \bar g _i (\bar t))
\label{CW}
\end{equation}
where $\bar \eta(\bar t)$ and $\bar g_i(\bar t)$ are defined by (\ref{rykarensdal}) and (\ref{carlotta}) with the $\bar \gamma_\phi$ and $\bar \beta_i$, and with a new running variable $\bar t$. Here we take:
\begin{equation}
\bar t=\log \frac \phi M
\label{ambercoin}
\end{equation}
 and thus $M(\bar t)=\phi$. This choice of $\bar t$ makes terms involving logarithmic factors $$\log \frac{\phi}{M(\bar t)}$$ disappear on the right-hand side of  (\ref{CW}). The  two solutions (\ref{FJSE}) and (\ref{CW}) of (\ref{oiseauxdefeu}) are equivalent.

The renormalization group equations (\ref{carlotta}) are solved in the leading and next-leading logarithmic approximation, where for $g_i^2=(\lambda, g^2, g'^2), i=1,2,3$:
\begin{eqnarray}&&
(g_i^2) ^{\{0\}}(t)=g_i^2+\sum _{m_1+m_2+m_3=n>1} a_{m_1, m_2, m_2}g_1^{2m_1}g_2^{2m_2}g_3^{2m_3}t^{n-1}
\label{doyoun}
\end{eqnarray}
where the coefficients $a_{m_1, m_2, m_3}$ are determined by (\ref{carlotta}) and the leading logarithmic approximation is indicated by a superscript $^{\{0\}}$.  In next-leading logarithmic order one gets, with the next-leading logarithmic approximation  indicated by a superscript $^{\{1\}}$, instead of (\ref{doyoun}):
\begin{eqnarray}&&
(g_i^2) ^{\{1\}}(t)=\sum _{m_1+m_2+m_3=n>1} b_{m_1, m_2, m_2}g_1^{2m_1}g_2^{2m_2}g_3^{2m_3}t^{n-2}
\label{kahyasi}
\end{eqnarray} 
with new coefficients $b_{m_1, m_2, m_3}$.

 For the  gauge parameter $\xi$ one also has a renormalization group equation with a solution  $\xi(t)$  that also depends on $\xi$, and $\xi(t)$ also has a leading and next-leading logarithmic approximation given by power series like (\ref{doyoun}) and (\ref{kahyasi}) (more details are given in the appendix).
In an expansion involving leading and next-leading logarithms one also  has the runnings Higgs boson field expectation value:
\begin{equation}
\phi^{\{0\}}(t)\simeq \phi \exp(-\int _0^tdt'\gamma_\phi^{\{0\}}(t')), \phi ^{\{1\}}(t)\simeq -\int _0^tdt'\gamma_\phi^{\{1\}}(t')\phi^{\{0\}}(t)
\label{balouch}
\end{equation}
where $\gamma _\phi$ is the anomalous dimension of the scalar field,   and details on $\gamma^{\{0\}}_\phi(t)$ and $\gamma^{\{1\}}_\phi(t)$  can be found in the appendix.

 (\ref{nickelcoin})    reduces to 
\begin{equation}
t\simeq \log \frac \phi M
\label{ohollywah}
\end{equation}
 in the leading logarithmic approximation, and in the next-leading logarithmic approximation one  gets:
\begin{equation}
 t\simeq  \log \frac{\phi }{ M}-\int _0^{\log \frac \phi M}dt'\gamma^{\{0\}} _\phi  (t').
\label{arason}
\end{equation}
Inserting here (\ref{ambercoin}) one gets (\ref{arason}) in the form:
\begin{equation}
t\simeq \bar t-\int _0^{\bar t}d t'\gamma_\phi ^{\{0\}}( t'), \ dt\simeq d\bar t(1-\gamma^{\{0\}}(\bar t)).
\label{carmilla}
\end{equation}

Combining  (\ref{balouch}) with (\ref{arason}) one finds in the next-leading logarithmic order:
\begin{equation}
\phi^{\{0\}}(t)+\phi^{\{1\}}(t)\simeq  (1-\int _0^{\log \frac \phi M}dt'\gamma_\phi^{\{1\}}(t')+\gamma^{\{0\}}(\log \frac \phi M)\int _0^{\log \frac \phi M}dt'\gamma_\phi^{\{0\}}(t'))\phi^{\{0\}}(\log \frac \phi M ).
\label{millarca}
\end{equation}
This equation should be compared with the next-leading logarithmic approximation of the  quantity $\bar \eta(\bar t)$ obtained from Coleman and Weinberg's solution (\ref{CW}):
\begin{equation}
\bar  \eta^{\{0\}}(\bar t)+\bar \eta ^{\{1\}}(\bar t)\simeq(1-\int _0^{\bar t}d\bar t'(\gamma^{\{1\}}(\bar t')-(\gamma^{\{0\}}(\bar t'))^2)) \exp(-\int _0^{\bar t}d\bar t'\gamma_\phi^{\{0\}}(\bar t')).
\label{sheridan}
\end{equation}
The discrepancy between (\ref{millarca}) and (\ref{sheridan}) is caused by the fact that the two integration variables $t$ and $\bar t$ are related by the transformation (\ref{carmilla}), which mixes different orders in the expansion in leading and next-leading logarithms, and it is removed by carrying out in   (\ref{balouch}) a change of integration variable  by (\ref{carmilla}).

Similarly one gets the quartic coupling constant $\lambda $ in the leading and next-leading logarithmic order:
\begin{eqnarray}&&
\lambda ^{\{0\}}(t)+\lambda^{\{1\}}(t) -\lambda\simeq \int _0^{ t}d  t' (\beta_\lambda^{\{0\}}(t')+\beta_\lambda^{\{1\}}(t'))
\nonumber\\&&
\simeq \int _0^{ \log \frac \phi M}d  t' (\beta_\lambda^{\{0\}}(t')+\beta_\lambda^{\{1\}}(t'))-\beta_\lambda^{\{0\}}(\log \frac \phi M) \int _0^{\log \frac \phi M}dt' \gamma_\phi ^{\{0\}}(t')
\label{frankenstein}
\end{eqnarray}
and:
\begin{equation}
\bar \lambda ^{\{0\}}(\bar t)+\bar \lambda^{\{1\}}(\bar t)-\lambda\simeq  \int _0^{\bar t}d  \bar t' (\beta_\lambda^{\{0\}}(\bar t')+\beta_\lambda^{\{1\}}(\bar t')-\gamma_\phi ^{\{0\}}(\bar t')\beta_\lambda^{\{0\}}(\bar t'))
\label{igor}
\end{equation}
where (\ref{frankenstein}) is converted into (\ref{igor}) through the transformation (\ref{carmilla}).

\section{Leading and next-leading logarithmic approximation of the effective potential}

The renormalization group improved potential is in the leading logarithmic approximation by (\ref{hedeprinsesse}):
\begin{equation}
V^{\{0\}}\simeq \frac 14 \lambda ^{\{0\}}(t)\phi^{\{0\}}(t)^4\mid _{t=\log \frac{\phi}{M}}
\label{austraat}
\end{equation}
 where $ \lambda ^{\{0\}}(t)$ and $\phi^{\{0\}}(t)$ only include leading logaritms. (\ref{austraat}) follows immediately from both the renormalization group equation solutions (\ref{FJSE}) and (\ref{CW}).
 Since the anomalous dimension, in contrast to the running coupling constant, is gauge parameter dependent, one gets from (\ref{austraat}):
\begin{eqnarray}&&
\xi \frac{\partial}{\partial\xi} V^{\{0\}}\simeq  -4\int_0^{t}dt'\xi \frac{\partial \gamma^{\{0\}}_\phi(t') }{\partial \xi}\mid _{t=\log \frac{\phi}{M}}V^{\{0\}}, 
\nonumber\\&&
\xi' \frac{\partial}{\partial\xi'} V^{\{0\}}\simeq  -4\int_0^{t}dt'\xi' \frac{\partial \gamma^{\{0\}}_\phi(t') }{\partial \xi'}\mid _{t=\log \frac{\phi}{M}}V^{\{0\}}.
\label{ygdrasil}
\end{eqnarray}
These relations  are  consistent with the partial differential equations  (\ref{identity}) in the context of the leading logarithms:
\begin{equation}
\xi \frac{\partial}{\partial\xi} V^{\{0\}}\simeq  -C^{\{0\}}\frac{\partial}{\partial \phi} V^{\{0\}}, \
\xi '\frac{\partial}{\partial\xi'} V^{\{0\}}\simeq  -C'^{\{0\}}\frac{\partial}{\partial \phi} V^{\{0\}}\label{robellino}
\end{equation}
 with:
\begin{eqnarray}&&
C^{\{0\}}=\phi\xi \frac{\partial}{\partial \xi}
\int _0^{\log \frac {\phi}{M}}dt'\gamma ^{\{0\}}_\phi(t'), \ \
\  C'^{\{0\}} =\phi \xi '\frac{\partial}{\partial \xi'}
\int _0^{\log \frac {\phi}{M}}dt'\gamma ^{\{0\}}_\phi(t')
\label{maegaard}
\end{eqnarray}
valid since the differentiation of the logarithms in (\ref{austraat}) convert leading logarithmic terms into next-leading logarithmic terms that are neglected in this approximation.

Including next-leading logarithms one gets from (\ref{hedeprinsesse}) and (\ref{mazanderan}) combined with (\ref{FJSE}) and using (\ref{arason}) the asymptotic effective potential:
\begin{eqnarray}&&
V^{\{0\}}+V^{\{1\}}\simeq \bigg (\frac 14 \lambda ^{\{0\}}(t)\phi^{\{0\}}(t)^4+\frac 14 \lambda ^{\{1\}}(t)\phi^{\{0\}}(t)^4+\lambda ^{\{0\}}(t)\phi^{\{0\}}(t)^3\phi^{\{1\}}(t)
\nonumber\\&&
+\frac 14 \Delta \lambda ^{\{0\}}(t)\phi^{\{0\}}(t)^4\bigg )\mid _{t=\log \frac{\phi}{M}-\int _0^{\log \frac \phi M}dt'\gamma^{\{0\}} _\phi  (t')}.
\label{danzig}
\end{eqnarray}
Here the coupling constants as well as the field  are taken only to the leading logarithmic approximation in the last   term, but in the first three terms both leading and next-leading logarithms are included.
This is because the expression $\Delta \lambda$ as seen from (\ref{ignoramus}) has two extra powers of $g$ or $ g'$  or one extra power of $\lambda$ in front
and hence the leading  logarithmic approximation of this term matches the
next-leading logarithmic approximation of the renormalization group improved tree potential.

Carrying out an expansion of the first term on the right-hand side of  (\ref{danzig}) one obtains  in this approximation:
\begin{eqnarray}&&
V^{\{1\}}\simeq \bigg (\frac 14 \lambda ^{\{1\}}(t)\phi^{\{0\}}(t)^4+\lambda ^{\{0\}}(t)\phi^{\{0\}}(t)^3\phi^{\{1\}}(t)
\nonumber\\&&
-\frac 14(\beta _\lambda ^{\{0\}}(t)-4\lambda ^{\{0\}}(t)\gamma _\phi^{\{0\}}(t) )\phi^{\{0\}}(t)^4\int _0^tdt'\gamma^{\{0\}}_\phi (t')
\nonumber\\&&
+\frac 14 \Delta \lambda ^{\{0\}}(t)\phi^{\{0\}}(t)^4\bigg )\mid _{t=\log \frac{\phi}{M}}.
\label{gdingen}
\end{eqnarray}
Using  the solution (\ref{CW}) of the renormalization group equation  one obtains a different result:
\begin{eqnarray}&&
V^{\{1\}}\simeq \bigg (\frac 14 \lambda ^{\{1\}}(t)\phi^{\{0\}}(t)^4+\lambda ^{\{0\}}(t)\phi^{\{0\}}(t)^3\phi^{\{1\}}(t)
\nonumber\\&&
- \frac 14\int _0^tdt' (\beta_\lambda ^{\{0\}}(t')
-4\lambda^{\{0\}}(t)\gamma_\phi ^{\{0\}}(t'))\gamma _\phi ^{\{0\}}(t')\phi^{\{0\}}(t)^4
\nonumber\\&&
+\frac 14 \Delta \lambda ^{\{0\}}(t)\phi^{\{0\}}(t)^4\bigg )\mid _{t=\log \frac{\phi}{M}}.
\label{kandis}
\end{eqnarray}

From (\ref{gdingen}) one gets, using (\ref{balouch}):
\begin{eqnarray}&&
\xi \frac{\partial}{\partial \xi} V^{\{1\}}\simeq - 4\int_0^{t}dt'\xi \frac{\partial \gamma^{\{0\}}_\phi(t') }{\partial \xi} \mid _{t=\log \frac{\phi}{M}}V^{\{1\}}
\nonumber\\&&
+\bigg(-4\int_0^{t}dt'\xi \frac{\partial \gamma^{\{1\}}_\phi(t')}{\partial \xi}
+4\xi \frac{\partial \gamma^{\{0\}} _\phi(t)}{\partial \xi}
\int _0^tdt'\gamma^{\{0\}}_\phi (t')
\nonumber\\&&
 -\int_0^tdt'\xi \frac{\partial \gamma ^{\{0\}}_\phi(t')}{\partial\xi}(\frac{\beta _\lambda ^{\{0\}}(t)}{\lambda ^{\{0\}}(t)}-4\gamma _\phi^{\{0\}}(t))
+\xi \frac{\partial }{\partial \xi}\frac{\Delta \lambda ^{\{0\}}(t)}{\lambda ^{\{0\}}(t)}\bigg )\mid _{t=\log \frac{\phi}{M}}V^{\{0\}}.
\nonumber\\&&
\label{hannaglawari}
\end{eqnarray}
(\ref{ygdrasil}) and (\ref{hannaglawari}) are combined with:
\begin{equation}
\phi\frac{\partial }{\partial \phi}V^{\{0\}}\simeq 4V^{\{0\}}+(\frac{\beta_\lambda ^{\{0\}}(t)}{\lambda ^{\{0\}}(t)}-4\gamma_\phi^{\{0\}}(t))\mid _{t=\log \frac{\phi}{M}}V^{\{0\}}
\label{karenjespersen}
\end{equation}
correct to the next-leading logarithmic approximation. Thus  (\ref{ygdrasil}) implies in this approximation:
\begin{equation}
\xi \frac{\partial}{\partial \xi}V^{\{0\}}\simeq -C^{\{0\}}\frac{\partial }{\partial \phi}V^{\{0\}}
+\int_0^{t}dt'\xi \frac{\partial \gamma^{\{0\}}_\phi(t') }{\partial \xi}(\frac{\beta_\lambda ^{\{0\}}(t)}{\lambda ^{\{0\}}(t)}-4\gamma_\phi^{\{0\}}(t))\mid _{t=\log \frac{\phi}{M}}V^{\{0\}}
\label{aleqahammond}
\end{equation}
and adding (\ref{hannaglawari}) and (\ref{aleqahammond}), using again (\ref{karenjespersen}), one obtains:
\begin{equation}
\xi \frac{\partial}{\partial \xi}(V^{\{0\}}+V^{\{1\}})
\simeq -C^{\{0\}}\frac{\partial }{\partial \phi}(V^{\{0\}}+V^{\{1\}})-C^{\{1\}}\frac{\partial }{\partial \phi}V^{\{0\}}
\label{sandynelson}
\end{equation}
 correct to the next-leading logarithmic approximation, with:
\begin{equation}
C^{\{1\}}=
\phi \bigg (\int_0^{t}dt'\xi \frac{\partial \gamma^{\{1\}}_\phi(t')}{\partial \xi}-\xi \frac{\partial\gamma^{\{0\}} _\phi(t)}{\partial \xi}
\int _0^tdt'\gamma_\phi^{\{0\}} (t')-\frac 14\xi \frac{\partial }{\partial \xi}\frac{\Delta \lambda ^{\{0\}}(t)}{\lambda ^{\{0\}}(t)}\bigg )\mid _{t=\log \frac{\phi}{M}}.
\label{willienelson}
\end{equation}
A similar construction leads to:
\begin{equation}
C'^{\{1\}}=
-\phi \bigg (\xi' \frac{\partial\gamma ^{\{0\}}_\phi(t)}{\partial \xi'}
\int _0^tdt'\gamma_\phi ^{\{0\}} (t') +\frac 14\xi '\frac{\partial }{\partial \xi'}\frac{\Delta \lambda ^{\{0\}}(t)}{\lambda ^{\{0\}}(t)}\bigg )\mid _{t=\log \frac{\phi}{M}}
\label{nelsonthegreat}
\end{equation}
where it should be kept in mind that $\gamma^{\{1\}}_\phi(t)$ does not depend on $\xi'$ since the anomalous dimension is independent of $\xi'$ at two-loop order, and at one-loop order it only depends on $\xi'$ in the renormalization group invariant combination $\xi'g'^2$ which only contributes to $\gamma_\phi^{\{0\}}(t)$.

Using (\ref{kandis}) as the next-leading logarithmivc approximation of the effective potential one gets instead of (\ref{sandynelson}):
\begin{equation}
\xi \frac{\partial}{\partial \xi}(V^{\{0\}}+V^{\{1\}})\simeq  -C^{\{0\}}\frac{\partial }{\partial \phi}(V^{\{0\}}+V^{\{1\}})-(C^{\{1\}}+\Delta C^{\{1\}})\frac{\partial }{\partial \phi}V^{\{0\}}
\label{alexander}
\end{equation}
where:
\begin{equation}
\Delta C^{\{1\}}=\phi \xi\frac{\partial }{\partial \xi} \bigg (\int_0^tdt' \gamma ^{\{0\}}_\phi(t')(\frac 14\frac{\beta _\lambda ^{\{0\}}(t')-\beta _\lambda ^{\{0\}}(t)}{\lambda ^{\{0\}}(t)}-(\gamma _\phi^{\{0\}}(t')-\gamma _\phi^{\{0\}}(t))
\bigg )\mid _{t=\log \frac{\phi}{M}}
\label{rokkakudo}
\end{equation}
with similar equations involving the gauge parameter $\xi'$.

\section {Field redefinition}

 All ingredients are now available to solve (\ref{Fukuda})-(\ref{Kugo}) in the context of the renormalization group improved effective potential in the leading and next-leading logarithmic approximation.

The leading logarithmic approximation of the effective potential is given by (\ref{austraat}). Here a solution of (\ref{Fukuda})-(\ref{Kugo})  is:
\begin{equation}
 \Phi ^{\{0\}}\simeq \phi \exp (-\int _0^{\log\frac{\phi}{M}}dt'(\gamma_\phi ^{\{0\}}(t')-\gamma_{\phi, 0} ^{\{0\}}(t')))
\label{lallahwatts}
\end{equation}
with:
\begin{equation}
\xi  \frac{\partial \Phi ^{\{0\}}}{\partial \xi}\simeq-C^{(0)}\frac{\Phi^{\{0\}}}{\phi}\simeq-C^{(0)}\frac{\partial \Phi ^{\{0\}}}{\partial \phi}
\label{macawally}
\end{equation}
where $C^{\{0\}}$ is given in (\ref{maegaard}), and with a similar equation for $\xi'$, and where $\gamma_{\phi, 0}$  denotes the value of the anomalous dimension $\gamma_\phi$  at the  specific values  $\xi_0$ and $\xi'_0$ of the gauge parameters.   Here was used that differentiation of the exponential of (\ref{ohollywah}) with respect to $\phi$ only gives a nonvanishing contribution in the next-leading logarithmic approximation. 

The  new running  variable  is:
\begin{equation}
\tau\simeq \log \frac{\Phi^{[0]}}{M}\simeq \log \frac \phi M-\int _0^{\log\frac{\phi}{M}}dt'(\gamma_\phi ^{\{0\}}(t')-\gamma_{\phi, 0} ^{\{0\}}(t'))
\label{coppercoin}
\end{equation}
according to (\ref{ohollywah}), using alao (\ref{lallahwatts}). The second term of (\ref{coppercoin}) is of next-leading logarithmic order, and thus:
\begin{equation} 
\tau \simeq t
\label{future}
\end{equation}
 at leading logarithmic order. Using the  anomalous dimension $\gamma _{\phi, 0}$ corresponding to the gauge parameters $\xi_0$ and $\xi'_0$ one next gets the new running field variable $\Phi^{\{0\}}(t)$ from (\ref{balouch}):
\begin{equation}
\Phi^{\{0\}}(t)=\Phi^{\{0\}}\exp(-\int _{0}^{t}dt' \gamma^{\{0\}} _{\phi, 0}(t') )
= \phi^{\{0\}}(t)
\label{issymckee}
\end{equation}
with $t\simeq \log \frac \phi M$, and with $\phi ^{\{0\}}(t)$  given in (\ref{balouch}).
Eliminating $\phi$ and introducing instead $\Phi ^{\{0\}}$ in (\ref{austraat}) one thus gets, correct to the leading  logarithmic approximation:
\begin{eqnarray}&&
\frac 14\lambda ^{\{0\}}(t)\phi^{\{0\}}(t)^4\mid _{t\simeq\log \frac \phi M}\simeq \frac 14 \lambda ^{\{0\}} (t)\Phi^{\{0\}}(t)^4\mid_{t \simeq \log \frac {\Phi^{\{0\}}}{ M}}
\label{parendal}
\end{eqnarray}
and this establishes the invariance of the effective potential under a change of the gauge parameters at leading logarithmic order in agreement with (\ref{Kyriakopoulos}).

At next-leading logarithmic order the solution of  (\ref{Fukuda})-(\ref{Kugo})  is instead of (\ref{lallahwatts}) for the renormalization group solution (\ref{FJSE}):
\begin{eqnarray}&&
 \Phi ^{\{0\}}+\Phi^{\{1\}}\simeq \Phi ^{\{0\}}\bigg(1+(\gamma^{\{0\}}_\phi(t)-\gamma^{\{0\}}_{\phi, 0}(t))\int _0^tdt' \gamma^{\{0\}}_\phi(t')
\nonumber\\&&
-\int _0^tdt'(\gamma^{\{1\}}_\phi (t')-\gamma^{\{1\}}_{\phi, 0}(t'))+\frac 14\frac{\Delta \lambda ^{\{0\}}(t)-\Delta \lambda_0 ^{\{0\}}(t)}{\lambda ^{\{0\}}(t)}\bigg)\mid _{t\simeq\log \frac{\phi}{M}}
\label{snoebrett}
\end{eqnarray}
since:
\begin{equation}
\xi \frac{\partial}{\partial \xi}(\Phi ^{\{0\}}+\Phi^{\{1\}})\simeq   -C^{\{0\}}\frac{\partial}{\partial \phi}(\Phi ^{\{0\}}+\Phi^{\{1\}})-C^{\{1\}}\frac{\partial}{\partial \phi}\Phi ^{\{0\}},
\label{shaunwhite}
\end{equation}
by (\ref{maegaard}) and (\ref{willienelson}),
with a similar equation for $\xi'$, and with:
\begin{equation}
\phi \frac{\partial}{\partial \phi}(\Phi ^{\{0\}}+\Phi^{\{1\}}) \simeq (1-(\gamma^{\{0\}}_\phi(t)-\gamma^{\{0\}}_{\phi, 0}(t)))\Phi ^{\{0\}}+\Phi^{\{1\}} 
\label{snobroed}
\end{equation}
by (\ref{lallahwatts}) and (\ref{snoebrett}), correct at next-leading logarithmic order.

It is next verified that the effective potential in the next-leading logarithmic approximation (\ref{gdingen}) is obtained also from the new field (\ref{snoebrett}). First it is shown that the new running variable still obeys (\ref{future}).  
From (\ref{lallahwatts}) and (\ref{snoebrett}) follows:
\begin{eqnarray}&&
\log \frac{\Phi^{\{0\}}+\Phi^{\{1\}}}{M}\simeq \log \frac \phi M
+\bigg((\gamma^{\{0\}}_\phi(t)-\gamma^{\{0\}}_{\phi, 0}(t))\int _0^tdt' \gamma^{\{0\}}_\phi(t')
\nonumber\\&&
-\int _0^tdt'( \gamma_\phi ^{\{0\}}(t')+\gamma_\phi^{\{1\}}(t')-\gamma_{\phi, 0} ^{\{0\}}(t')-\gamma^{\{1\}}_{\phi ,0}(t'))
+\frac 14\frac{\Delta \lambda ^{\{0\}}(t)-\Delta \lambda _0^{\{0\}}(t)}{\lambda ^{\{0\}}(t)}\bigg)\mid _{t\simeq\log \frac{\phi}{M}}
\nonumber\\&&
\label{lyphard}
\end{eqnarray}
correct in the next-leading logarithmic approximation, and thus the new running variable $\tau $ replacing $t$ is by (\ref{arason}):
\begin{eqnarray}&&
\tau \simeq  \log \frac{\Phi^{\{0\}}+\Phi^{\{1\}}}{M}-\int _0^{\log \frac{\Phi^{\{0\}+\Phi^{\{1\}}}}{M}}dt' \gamma _{\phi, 0}^{\{0\}}(t')
\nonumber\\&&
\simeq  t
-\int _{\log \frac \phi M}^{\log \frac{\Phi^{\{0\}}++\Phi^{\{1\}}}{M}}dt' \gamma _{\phi, 0}^{\{0\}}(t')
+\bigg((\gamma^{\{0\}}_\phi(t)-\gamma^{\{0\}}_{\phi, 0}(t))\int _0^tdt' \gamma^{\{0\}}_\phi(t')
\nonumber\\&&
-\int _0^tdt'( \gamma_\phi^{\{1\}}(t')-\gamma^{\{1\}}_{\phi ,0} (t'))
+\frac 14\frac{\Delta \lambda ^{\{0\}}(t)-\Delta \lambda_0 ^{\{0\}}(t)}{\lambda ^{\{0\}}(t)}\bigg)\mid _{t\simeq\log \frac{\phi}{M}}
\label{asparagus}
\end{eqnarray}
with the quantity $t$ given by (\ref{arason}), and  the  other terms on the right-hand side  are only nonvanishing at next-next-leading logarithmic order, and so (\ref{future})
 holds true also at next-leading logarithmic order.

The running field variable is thus instead of (\ref{issymckee}), keeping in mind that  the anomalous dimension $\gamma _{\phi, 0}$  should be used:
\begin{equation}
(\Phi^{\{0\}}+\Phi^{\{1\}})(t)=(\Phi^{\{0\}}+\Phi^{\{1\}})\exp(-\int _{0}^{t}dt' (\gamma^{\{0\}} _{\phi, 0}(t')+\gamma^{\{1\}}_{\phi, 0}(t')) )
\label{gamborg}
\end{equation}
where $t$ is given by (\ref{arason}), with:
\begin{eqnarray}&&
\exp(-\int _{0}^{t}dt' (\gamma^{\{0\}} _{\phi, 0}(t')+\gamma^{\{1\}}_{\phi, 0}(t')) )\mid_{t=\log \frac \phi M-\int _0^{\log \frac \phi M}dt'\gamma^{\{0\}} _\phi  (t')}
\nonumber\\&&
\simeq\exp(-\int _{0}^{t}dt' \gamma^{\{0\}} _{\phi, 0}(t'))
\bigg(1+\gamma^{\{0\}}_{\phi, 0}(t)\int _0^tdt'\gamma^{\{0\}}_\phi(t')
-\int _{0}^{t}dt' \gamma^{\{1\}} _{\phi, 0}(t')\bigg)\mid _{t\simeq\log \frac{\phi}{M}}.
\nonumber\\&&
\label{klorofyl}
\end{eqnarray}
 Inserting  (\ref{lallahwatts}), (\ref{snoebrett}) and (\ref{klorofyl}) into (\ref{gamborg}) one then finds:
\begin{equation}
(\Phi^{\{0\}}+\Phi^{\{1\}})(t)
\simeq \phi^{\{0\}}(t)(1+\frac 14\frac{\Delta \lambda ^{\{0\}}(t)-\Delta \lambda_0 ^{\{0\}}(t)}{\lambda ^{\{0\}}(t)})+\phi^{\{1\}}(t)
\label{langtbortistan}
\end{equation}
with $t$  again  given by (\ref{arason}),
and  (\ref{danzig}) can by (\ref{langtbortistan}) be tested for invariance under a change of the gauge parameters:
\begin{eqnarray}&&
 \frac 14  (\lambda^{\{0\}}+\lambda ^{\{1\}}) (t)(\Phi  ^{\{0\}}+\Phi^{\{1\}})(t)^4 +\frac 14\Delta \lambda_0 ^{\{0\}}(t)\Phi^{\{0\}}(t)^4
\nonumber\\&&
\simeq \frac 14(\lambda^{\{0\}}+\lambda ^{\{1\}})(t)  (\phi^{\{0\}}+\phi^{\{1\}})(t)^4+\frac 14\Delta\lambda ^{\{0\}}(t)\phi ^{\{0\}}(t)^4
\label{ballesteros}
\end{eqnarray}
which completes the proof  that the renormalization group improved effective potential is gauge parameter independent in the next-leading logarithmic approximation in the context of the renormalization group solution (\ref{FJSE}), again in agreement with (\ref{Kyriakopoulos}).

Using  (\ref{kandis}) as the next-leading approximation of the effective potential the solution of (\ref{Fukuda})-(\ref{Kugo}) contains at next-leading logarithmic order  in addition to (\ref{snoebrett}), as seen from (\ref{rokkakudo}):
\begin{eqnarray}&&
\Delta \Phi^{\{1\}}\simeq 
-\Phi ^{\{0\}}\bigg(\int_0^tdt' \gamma ^{\{0\}}_\phi(t')(\frac 14\frac{\beta _\lambda ^{\{0\}}(t')-\beta _\lambda ^{\{0\}}(t)}{\lambda ^{\{0\}}(t)}-(\gamma _\phi^{\{0\}}(t')-\gamma _\phi^{\{0\}}(t)))
\nonumber\\&&
-\int_0^tdt' \gamma ^{\{0\}}_{\phi, 0}(t')(\frac 14\frac{\beta _\lambda ^{\{0\}}(t')-\beta _\lambda ^{\{0\}}(t)}{\lambda ^{\{0\}}(t)}-(\gamma _{\phi, 0}^{\{0\}}(t')-\gamma _{\phi, 0}^{\{0\}}(t)))
\bigg)\mid _{t\simeq\log \frac{\phi}{M}}
\label{eaudecologne}
\end{eqnarray}
with:
\begin{equation}
\xi \frac{\partial}{\partial \xi}(\Phi ^{\{0\}}+\Phi^{\{1\}}+\Delta \Phi^{\{1\}})\simeq   -C^{\{0\}}\frac{\partial}{\partial \phi}(\Phi ^{\{0\}}+\Phi^{\{1\}}+\Delta \Phi^{\{1\}})-(C^{\{1\}}+\Delta C^{\{1\}})\frac{\partial}{\partial \phi}\Phi ^{\{0\}},
\label{schaumwein}
\end{equation}
using again (\ref{snobroed}) and again with a similar equation for $\xi'$.

The running variable  is in this case by (\ref{ambercoin}):
\begin{equation}
\bar \tau =\log \frac{\Phi^{\{0\}}+\Phi^{\{1\}}+\Delta \Phi^{\{1\}}}{M}
\label{sorteslyngel}
\end{equation}
which  in the next-leading logarithmic approximation reduces to (\ref{coppercoin}), and thus:
\begin{equation}
\lambda^{\{0\}}(\bar \tau)\simeq(\lambda ^{\{0\}}(t)-\beta_\lambda^{\{0\}}(t)\int _0^{t}dt'(\gamma_\phi ^{\{0\}}(t')-\gamma_{\phi, 0} ^{\{0\}}(t')))\mid _{t=\log \frac \phi M}.
\label{dimeadozen}
\end{equation}
For $\Phi^{\{1\}}(\bar \tau)$ and $\Delta \Phi^{\{1\}}(\bar \tau)$ it is sufficient to take $\bar\tau \simeq \log \frac \phi M$ at next-leading logarithmic order. One thus finds by (\ref{balouch}), (\ref{coppercoin}), (\ref{issymckee}), (\ref{snoebrett}) and (\ref{eaudecologne}) the running field variable in this case:
\begin{eqnarray}&&
(\Phi^{\{0\}}+\Phi^{\{1\}}+\Delta \Phi^{\{1\}})(\bar \tau)
\nonumber\\&&
 = (\Phi^{\{0\}}+\Phi^{\{1\}}+\Delta \Phi^{\{1\}})
\exp(-\int _{0}^{\bar \tau }dt' (\gamma^{\{0\}} _{\phi, 0}(t')+\gamma^{\{1\}} _{\phi, 0}(t') ))
\nonumber\\&&
\simeq \bigg(\phi^{\{1\}}(t)+\phi ^{\{0\}}(t)
\bigg(1
+\int _0^tdt'((\gamma _\phi^{\{0\}}(t'))^2-(\gamma _{\phi, 0}^{\{0\}}(t'))^2)+\frac 14\frac{\Delta \lambda ^{\{0\}}(t)-\Delta \lambda _0^{\{0\}}(t)}{\lambda ^{\{0\}}(t)}
\nonumber\\&&
-\frac 14\int_0^tdt'( \gamma ^{\{0\}}_\phi(t')-\gamma ^{\{0\}}_{\phi, 0}(t'))\frac{\beta _\lambda ^{\{0\}}(t')-\beta _\lambda ^{\{0\}}(t)}{\lambda ^{\{0\}}(t)}\bigg)\bigg)\mid _{t=\log \frac \phi M}
\label{tmutorakan}
\end{eqnarray}
and therefore, using again (\ref{issymckee}):
\begin{eqnarray}&&
\frac 14 \lambda^{\{0\}}(\bar \tau )\Phi^{\{0\}}(\bar \tau)^4+ \frac 14 \lambda ^{\{1\}}(t)\Phi^{\{0\}}(t)^4+\lambda^{\{0\}}(t)\Phi^{\{0\}}(t)^3(\Phi^{\{1\}}+\Delta \Phi^{\{1\}})(t)
\nonumber\\&&
- \frac 14\int _0^tdt' (\beta_\lambda ^{\{0\}}(t')
-4\lambda^{\{0\}}(t)\gamma_{\phi, 0} ^{\{0\}}(t'))\gamma _{\phi, 0} ^{\{0\}}(t')\Phi^{\{0\}}(t)^4
+\frac 14 \Delta \lambda_0 ^{\{0\}}(t)\Phi^{\{0\}}(t)^4
\nonumber\\&&
\simeq \frac 14\lambda^{\{0\}}(t)\phi^{\{0\}}(t)^4+\frac 14 \lambda ^{\{1\}}(t)\phi^{\{0\}}(t)^4+\lambda ^{\{0\}}(t)\phi^{\{0\}}(t)^3\phi^{\{1\}}(t)
\nonumber\\&&
- \frac 14\int _0^tdt' (\beta_\lambda ^{\{0\}}(t')
-4\lambda^{\{0\}}(t)\gamma_\phi ^{\{0\}}(t'))\gamma _\phi ^{\{0\}}(t')\phi^{\{0\}}(t)^4
+\frac 14 \Delta \lambda ^{\{0\}}(t)\phi^{\{0\}}(t)^4
\nonumber\\&&
\label{salamander}
\end{eqnarray}
valid in next-leading logarithmic order, with $t=\log \frac \phi M$ and $\bar \tau$ given by (\ref{sorteslyngel}). 
 (\ref{salamander})  verifies (\ref{Kyriakopoulos}) for this case also. This proves according to (\ref{kandis})  that the effective potential is unchanged under gauge parameter changes for the renormalization group solution (\ref{CW}) in the next-leading logarithmic order too.

\section{Conclusion and comments}

It has been demonstrated above that the equation (\ref{identity}) allows a redefinition of the field variable of the effective potential in the $SU(2)\otimes U(1)$ model of electroweak unification in the leading and next-leading logarithmic approximation  for both the renormalization group equation solutions (\ref{FJSE}) and (\ref{CW}). As a result of the redefinition, the arbitary gauge parameters $(\xi, \xi')$ are eliminated in terms of fixed  gauge parameters $(\xi_0, \xi'_0)$ of some preferred gauge such as the Landau gauge. The redefinition is achieved by solving (\ref{Fukuda})-(\ref{Kugo}) to the required accuracy, and the solutions are (\ref{lallahwatts}) in the leading logarithmic approximation and (\ref{snoebrett}) and (\ref{eaudecologne}) in the next-leading logarithmic approximation. Remarkably,  the construction in the leading logarithmic approximation only involves the anomalous dimension of the Higgs boson field, while in the next leading logarithmic approximation the quantity $\Delta \lambda$ defined in (\ref{ignoramus}) plays a crucial role.  

As mentioned in the introduction,  (\ref{lallahwatts}), (\ref{snoebrett})  and (\ref{eaudecologne}) define mappings  of the field variable $\phi$, and the inverse mappings are obtained by interchanging $(\xi, \xi')$ and $(\xi_0, \xi'_0)$ in   (\ref{lallahwatts}), (\ref{snoebrett})  and (\ref{eaudecologne}); this is also easily  shown directly and represents a consistency check on the solutions of (\ref{Fukuda}) and (\ref{Kugo}) represented by (\ref{lallahwatts}), (\ref{snoebrett})  and (\ref{eaudecologne}).  This observation makes it possible to  obtain the field variable in an arbitrary gauge from the field variable in the Landau gauge as used in \cite{Cabibbo}, \cite{Casas}, \cite{Degrassi} such that the effective potential is invariant: one substitutes  instead of the field variable $\phi$ the solutions   (\ref{lallahwatts}), (\ref{snoebrett})  and (\ref{eaudecologne})  determined above, with the replacements  $(\xi, \xi')\rightarrow (0, 0)$ and  $(\xi_0, \xi'_0)\rightarrow (\xi, \xi')$, since one is now dealing with the inverse  mapping of the field variable.  The gauge parameter depencence of the   instability scale studied numerically in \cite{Luzio} can be found analytically this way. 

The conditions (\ref{nickelcoin}) and (\ref{ambercoin}) were imposed on the renormalization group running variables in order to obtain well-defined approximation schemes. They may not be the optimal choices of  running variables from the point of view of numerical accuracy. Also it was pointed out that the two solutions (\ref{FJSE}) and (\ref{CW}), though equivalent, have different expansions  in this approximation scheme, and thus one may give more precise estimates than the other one. Similarly, different gauge choices may lead to different degrees of accuracy, though they are formally equivalent.

{\bf Acknowledgements.} I wish to thank Dr. A. Fettouhi and Dr. K. Boye for taking part in the early stages of the investigation presented here. Also I am grateful to Professor Peter van Nieuwenhuizen at the C. N. Yang Institute of Theoretical Physics, State University of New York at Stony Brook, where this work was initiated some years ago, for his hospitality. Finally I  would like to express my gratitude to Professor J.-M.Fr\'{e}re, Professor P.M.Lavrov and Professor O.M.Del Cima for   drawing my attention to their work, and to a referee for his questions and comments.

\appendix

\section{Renormalization group functions}

$g_i$ represents  the  couplings of the theory 
(including the dimensionless gauge fixing parameters $\xi, \xi'$) that obey the renormalization group equations (\ref{carlotta}).  Solving these equations with the one-loop $\beta$-functions leads to the leading logarithmic approximation of the running coupling constants, and including also two-loop $\beta$-functions one obtains the next-leading logarithmic approximation.  The $\beta$-functions  for coupling constants are gauge parameter independent, and up to two-loop order they can be seen from \cite{Casas}, \cite{Degrassi}, \cite{Andreassen}.

Anomalous dimensions are, in contrast to coupling constant $\beta$-functions,   dependent on the dimensionless gauge fixing parameters. The anomalous dimension of the $W$-field in the electroweak theory is at one-loop order, keeping only  bosonic contributions:
\begin{equation} 
\gamma^{[1]} _W \simeq -\frac{1}{16\pi ^2}(\frac {25}{6}-\xi)g^2
\label{angela}
\end{equation}
and the function $\beta _\xi$ for the gauge parameter $\xi$ is in general given by:
\begin{equation}
\beta_\xi=-2\xi \gamma _W.
\label{lateran}
\end{equation}
Solving (\ref{carlotta}) for the gauge parameter $\xi$  with the $\beta $-function (\ref{lateran}) one  gets, using the one-loop value of  $\beta_\xi$, the running gauge parameter in the leading logarithmic approximation. The  quantity $\xi' g'^2$ is  renormalization group invariant.
The anomalous dimension of the scalar field is at one-loop order, disregarding fermions:
 \begin{equation}
\gamma^{[1]}_\phi \simeq 
  -\frac{1}{16\pi^2}(\frac 34(3g^2+g'^2)-\frac 34\xi g^2-\frac 14\xi' g'^2).
\label{annyschlemm}
\end{equation}
Inserting here the leading logarithmic approximations of the coupling  constants and gauge fixing parameters one obtains the leading logarithmic approximation of the anomalous dimension, the function $\gamma^{\{0\}}_\phi(t)$ used in the text.

The anomalous dimension of the $W$-field is at two-loop order, keeping only terms involving the coupling constant $g$ (gauge parameter dependence only occurs in these terms):
\begin{equation}
\gamma^{[2]}_W\simeq (\frac{1}{16\pi^2})^2(-\frac{231}{8}+\frac{11}{2}\xi+\xi^2)g^4
\label{hasdrubal}
\end{equation}
where the value in pure $SU(2)$ Yang-Mills theory is found from \cite{Tarasov} and the scalar field contribution from \cite{Machacek}.
$\beta_\xi$ is determined by (\ref{lateran}) also at two-loop order, and solving  (\ref{carlotta}) in this case one obtains the  running gauge parameter in the next-leading logarithmic approximation. 
The two-loop anomalous dimension of the scalar field $\gamma ^{[2]}_\phi$ is also found from
\cite{Machacek}. The part proportional to $g^4$ is, neglecting again the fermion contribution:
\begin{eqnarray}&&
\gamma ^{[2]}_\phi\simeq (\frac{1}{16\pi^2})^2(-\frac{511}{32}+3\xi+\frac 38\xi^2)g^4.
\label{matuchek}
\end{eqnarray}
$\gamma^{[2]}_\phi$ has no dependence on $\xi'$. Inserting $g^2$ and $\xi$ in  leading logarithmic order into (\ref{matuchek}) and  in next-leading logarithmic order into (\ref{annyschlemm}) the sum is the function $\gamma^{\{1\}}_\phi(t)$ (the result is only complete in the gauge parameter dependent part).

\end{document}